\documentclass[12pt]{amsproc}
\usepackage{amssymb}
\usepackage{amscd}
\usepackage{amsmath}

\theoremstyle{plain}

\newtheorem{definition}{Definition}

\newtheorem{lemma}{Lemma}

\newtheorem{remark}{Remark}

\newtheorem{theorem}{Theorem}
\numberwithin{equation}{section}

\begin{document}

\title[Weyl functions for Jacobi matrices]{On the relationship between\\ Weyl functions of \\Jacobi matrices
and response vectors for \\special dynamical systems with \\discrete
time}

\author{ A. S. Mikhaylov}
\address{St. Petersburg   Department   of   V.A. Steklov    Institute   of   Mathematics
of   the   Russian   Academy   of   Sciences, 7, Fontanka, 191023
St. Petersburg, Russia and Saint Petersburg State University,
St.Petersburg State University, 7/9 Universitetskaya nab., St.
Petersburg, 199034 Russia.} \email{mikhaylov@pdmi.ras.ru}

\author{ V. S. Mikhaylov}
\address{St.Petersburg   Department   of   V.A.Steklov    Institute   of   Mathematics
of   the   Russian   Academy   of   Sciences, 7, Fontanka, 191023
St. Petersburg, Russia and Saint Petersburg State University,
St.Petersburg State University, 7/9 Universitetskaya nab., St.
Petersburg, 199034 Russia.} \email{vsmikhaylov@pdmi.ras.ru}

\author{ S. A. Simonov}
\address{St.Petersburg   Department   of   V.A.Steklov    Institute   of   Mathematics
of   the   Russian   Academy   of   Sciences, 7, Fontanka, 191023
St. Petersburg, Russia and Saint Petersburg State University,
St.Petersburg State University, 7/9 Universitetskaya nab., St.
Petersburg, 199034 Russia.} \email{sergey.a.simonov@gmail.com}


\maketitle

\begin{abstract} We derive special representation for Weyl functions for finite
and semi-infinite Jacobi matrices with bounded entries based on a
relationship between spectral problem for Jacobi matrices and
initial-boundary value problem for auxiliary dynamical systems
with the discrete time for Jacobi matrices.
\end{abstract}

\section{Introduction.}

Given a sequence of real numbers $(b_1, b_2,\ldots )$ and
a sequence of positive numbers $(a_1, a_2,\ldots )$ such that
$\sup_{n\geqslant 1}\{a_n,|b_n|\}<\infty$, we consider an
operator $H$ corresponding to a semi-infinite Jacobi matrix,
defined on $l^2(\mathbb N)\ni \psi=(\psi_1, \psi_2, \ldots)$,
given by
\begin{align*}
(Hu)_n&=a_{n-1}\psi_{n-1}+b_n\psi_n+a_n\psi_{n+1},\quad
n\geqslant 2,\\
(Hu)_1&=b_1\psi_1+a_1\psi_2,\quad n=1.
\end{align*}

For a fixed $N\in \mathbb{N}$ we consider a finite Jacobi matrix:
\begin{equation*}
H^N=\begin{pmatrix}
b_1 & a_1 & 0 &\cdots &0&0\\
a_1& b_2 & a_2 &\cdots &0&0 \\
0& a_2 & b_3 & \cdots &0&0 \\
\vdots & \vdots & \vdots & \ddots & \vdots&\vdots\\
0 &0&0 &\cdots & b_{N-1}&a_{N-1}\\
0&0&0&\cdots&a_{N-1} & b_N
\end{pmatrix}.
\end{equation*}

The Weyl $m-$functions for $H$ and $H^N$ are defined by
\begin{eqnarray*}
m(\lambda)=\left( (H-\lambda)^{-1}e_1,e_1\right),\\
m^N(\lambda)=\left((H^N-\lambda)^{-1}e_1,e_1\right),
\end{eqnarray*}
$e_1=(1,0,0,\ldots)^T$.  By the spectral theorem, this definition is equivalent to the following:
\begin{equation*}
m(\lambda)=\int_{\mathbb{R}}\frac{d\rho(z)}{z-\lambda},
\end{equation*}
where $d\rho=d\left(E_He_1,e_1\right)$ and
$E_H$ is the (projection-valued) spectral measure of the self-adjoint operator $H$. The latter makes Weyl
function an important object in spectral and inverse spectral
theory of discrete and continuous one-dimensional systems
\cite{Ahiez,SG,T,BS1,AMR}.

Let us denote $\mathbb{N}_0:=\{0\}\cup \mathbb{N}$. For the same
sequences $(b_1, b_2,\ldots )$, $(a_1, a_2,\ldots )$, and an
additional parameter $a_0$, which for convenience we set equal to
one, we consider two dynamical systems with the discrete time. The
first system corresponds to semi-infinite case:
\begin{equation}
\label{Jacobi_dyn} \left\{
\begin{array}l
u_{n,t+1}+u_{n,t-1}-a_nu_{n+1,t}-a_{n-1}u_{n-1,t}-b_nu_{n,t}=0,\quad n,t\in \mathbb{N},\\
u_{n,-1}=u_{n,0}=0,\quad n\in \mathbb{N}, \\
u_{0,t}=f_t,\quad t\in \mathbb{N}_0,
\end{array}\right.
\end{equation}
and is a natural discrete analog of an initial-boundary value
problem for the wave equation on a half-line. The dynamic inverse
problem for such systems  was considered in \cite{AM,BM}. In
(\ref{Jacobi_dyn}) the real  sequence $f=(f_0,f_1,\ldots)$ is
interpreted as  a \emph{boundary control}. We denote the solution
to (\ref{Jacobi_dyn}) by $u^f_{n,t}$. With (\ref{Jacobi_dyn}) we
associate the \emph{response  operator} $R:
\mathbb{R}^\infty(\mathbb{N}_0)\mapsto
\mathbb{R}^\infty(\mathbb{N})$, acting by the rule:
\begin{equation}
\label{resp_infinit} \left(R f\right)_t:=u^f_{1,t},\quad
t=1,2,\ldots.
\end{equation}
The operator $R$ is a discrete analog of a dynamic
Dirichlet-to-Neumann operator, which is a classical object in
the theory of inverse dynamic problems \cite{B07,B01JII,KKL,I}.

The second system which corresponds to a finite case,
\begin{equation}
\label{Jacobi_dyn_int} \left\{
\begin{array}l
v_{n,t+1}+v_{n,t-1}-a_nv_{n+1,t}-a_{n-1}v_{n-1,t}-b_nv_{n,t}=0,\, t\in \mathbb{N}, n\in 1\ldots N,\\
v_{n,-1}=v_{n,0}=0,\quad n=1,2,\ldots,N+1, \\
v_{0,t}=g_t,\quad v_{N+1,t}=0,\quad t\in \mathbb{N}_0,
\end{array}\right.
\end{equation}
is a discrete analog of  an initial-boundary value problem for dynamical systems governed by the wave equation on a finite
interval. The solution to (\ref{Jacobi_dyn_int}) is denoted by
$v^g_{n,t}$. The real sequence $g=(g_0,g_1,\ldots)$ is a
\emph{boundary control}. With (\ref{Jacobi_dyn_int}) we associate
the \emph{response operators} $R_N:  \mathbb R^\infty(\mathbb{N}_0)\mapsto
\mathbb{R}^\infty(\mathbb{N})$ acting by the rule:
\begin{equation}
\label{resp_finit} \left(R_N f\right)_t:=v^g_{1,t},\quad t=1,2,\ldots
\end{equation}

It was shown in \cite{MM,MM1} that operators $R,\,R^N$ are
determined by their kernels, so-called \emph{response vectors}:
$\left(r_0,r_1,r_2,\ldots\right)$,
$\left(r_0^N,r_1^n,r_2^N,\ldots\right)$. We will establish the
relationship between  Weyl functions $m(\lambda)$, $m^N(\lambda)$
and response vectors. Such a relationship gives a new way of
calculation of Weyl functions, and at the same time it is
important in the inverse spectral theory: see
\cite{BS1,AMR,AM,MM2,MM3} for the case of  the Schr\"odinger
operator on a half-line. The main result of the paper is the
following.
\begin{theorem}
\label{Theor_main}
If the coefficients in the semi-infinite $H$ or
finite $H^N$ Jacobi matrix operators satisfy $\sup_{n\geqslant
1}\{a_n,|b_n|\}\leqslant B$, then the Weyl functions $m,$ $m^N$
admit the representations
\begin{eqnarray}
m(\lambda)=-\sum_{t=0}^\infty z^tr_t,\label{M_inf}\\
m^N(\lambda)=-\sum_{t=0}^\infty z^tr^N_t\label{M_fin}
\end{eqnarray}
in terms of response vectors (kernels of dynamic response
operators associated to (\ref{Jacobi_dyn}),
(\ref{Jacobi_dyn_int})): $\left(r_0,r_1,r_2,\ldots\right)$,
$\left(r_0^N,r_1^n,r_2^N,\ldots\right)$,  where the variables
$\lambda$ and $z$ are related by
\begin{equation*}
\lambda=z+\frac1z, \quad
z=\frac{\lambda-\sqrt{\lambda^2-4}}2=\frac{\lambda-i\sqrt{4-\lambda^2}}2.
\end{equation*}
These representations hold for $\lambda\in D$, where $D\subset
\mathbb{C}_+$  is defined as follows. Let $R:=3B+1$, then
\begin{eqnarray}
\label{D_reg} D:=\left\{x+iy\,\Big|\, x>
\left(R+\frac{1}{R}\right)\cos{\phi},\right.\\
\left.y>0,\,
y>\left(\frac{1}{R}-R\right)\sin{\phi},\,\phi\in(\pi,2\pi)\right\}.\notag
\end{eqnarray}
\end{theorem}

Our approach to Weyl function was stimulated by works on the
Boundary Control method \cite{B07,B17,B01JII} for dynamic inverse
problems. For the first time the relationship between Weyl
functions and auxiliary dynamical systems was established in
\cite{AMR}. The inverse problem of recovering a Jacobi matrix from
the response operator (i.e., from the response vector) was solved
in \cite{MM,MM1} by the Boundary Control method. The inverse
problem of recovering a Jacobi matrix from the Weyl function was
studied in \cite{SG}. Now, using Theorem \ref{Theor_main} one can
solve the second inverse problem by the following procedure: first
one needs to recover the response vector from $m(\lambda)$ or
$m^N(\lambda)$, which can be done by observing that in the
variable $z$ formulas (\ref{M_inf}), (\ref{M_fin}) are Taylor
expansions at zero, and then to use the dynamic method described
in \cite{MM1} to recover $a_i,\,b_i$, $i\geqslant 1$. We note that
for the first time the Boundary Control method was applied to
discrete dynamical systems in \cite{ABu,ABuN} in connection with
spectral estimation problem.

The paper is organized as follows. In the second section we
provide all necessary information on dynamical systems with
discrete time corresponding to Jacobi matrices. In the third
section we give  an (equivalent) definition of  the Weyl function
which is more relevant to our approach, introduce the ``Fourier
transform'' associated with the discrete Schr\"odinger operator
with zero potential, derive representation formulas for Weyl
functions $m^N$, $m$, and prove Theorem \ref{Theor_main}.

\section{Dynamical systems with discrete time corresponding to finite and semi-infinite Jacobi matrices.}

In this section we provide some results obtained in \cite{MM,MM1}.
We denote by $\mathcal{F}^\infty$ the \emph{outer space} of the
system (\ref{Jacobi_dyn}), the space of controls:
$\mathcal{F}^\infty:=\mathbb{R}^\infty$,  with $f=(f_0,f_1,f_2,\ldots)\in
\mathcal{F}^\infty$.
\begin{lemma}
The solution to (\ref{Jacobi_dyn}) admits the representation
\begin{equation}
\label{Jac_sol_rep}
u^f_{n,t}=\left\{\begin{array}l\prod\limits_{k=0}^{n-1}
a_kf_{t-n}+\sum\limits_{s=n}^{t-1}w_{n,s}f_{t-s-1},\quad n\leqslant
t,\,\, n,t\in
\mathbb{N},\\
0,\quad  n>t,\,\,n,t\in \mathbb{N},
\end{array}
\right.
\end{equation}
where $w_{n,s}$  is the solution of the  following Goursat problem
\begin{equation}
\label{Goursat} \left\{
\begin{array}l
w_{n,s+1}+w_{n,s-1}-a_nw_{n+1,s}-a_{n-1}w_{n-1,s}-b_nw_{n,s}=\\
=-\delta_{s,n}(1-a_n^2)\prod\limits_{k=0}^{n-1}a_k,\,n,s\in \mathbb{N}, \,\,s>n,\\
w_{n,n}-b_n\prod\limits_{k=0}^{n-1}a_k-a_{n-1}w_{n-1,n-1}=0,\quad n\in \mathbb{N},\\
w_{0,t}=0,\quad t\in \mathbb{N}_0.
\end{array}
\right.
\end{equation}
\end{lemma}
This representation formula is an analog  of the Duhamel formula
for  the wave equation with potential on a half-line \cite{AM}.

\begin{definition}
For $f,g\in \mathcal{F}^\infty$ we define the convolution
$c=f*g\in \mathcal{F}^\infty$ by the formula
\begin{equation*}
c_t=\sum_{s=0}^{t}f_sg_{t-s},\quad t\in  \mathbb{N}_0.
\end{equation*}
\end{definition}
In accordance with (\ref{Jac_sol_rep}) and  the definition of  the response
operator (\ref{resp_infinit}) (we  assume here that $a_0=1$):
\begin{eqnarray}
\label{R_def} \left(Rf\right)_t=u^f_{1,t}=f_{t-1}+\sum_{s=1}^{t-1}
w_{1,s}f_{t-1-s}
\quad t=1,2,\ldots,\\
\notag \left(R^Tf\right)=r*f_{\cdot-1},
\end{eqnarray}
where $r=(r_0,r_1,\ldots,r_{T-1})=(1,w_{1,1},w_{1,2},\ldots )$, is
 the \emph{response vector}, the convolution kernel of  the response
operator. Note that choosing the special control
$f=\delta=(1,0,0,\ldots)$, we can write the solution
(\ref{Jac_sol_rep}) as
\begin{equation}
\label{Duam1}
u^f_{n,t}=u^{\delta}_{n,\cdot}*f_{\cdot-1},
\end{equation}
and the response vector can be determined as
\begin{equation}
\label{con11} r_{t-1}=\left(R\delta\right)_t=u^\delta_{1,t}\quad
t\in \mathbb{N}.
\end{equation}

Similarly, for the system (\ref{Jacobi_dyn_int}) the solution
$v^g$ admits the representation in terms of convolution with a
special solution corresponding to control $\delta$:
\begin{equation}
\label{Duam2}
v^g_{n,t}=v^{\delta}_{n,\cdot}*g_{\cdot-1},
\end{equation}
and the response vector is given by
\begin{equation}
\label{con1}
r_{t-1}^{N}=\left(R_{N}\delta\right)_t=v^\delta_{1,t},\quad t\in
\mathbb{N}.
\end{equation}
\begin{remark}
Formulas (\ref{Duam1}), (\ref{Duam2}) are discrete analog of
classical Duhamel representation formula. They can be checked by
the direct calculations (see also \cite{MM,MM1}).
\end{remark}

\section{Weyl function representation}
 In this section we provide necessary information about the Weyl function and orthogonal
 polynomials associated with the Jacobi matrix, discrete Fourier transform related to the unperturbed
 discrete Schr\"odinger operator that we use for the variable $t$ of systems \eqref{Jacobi_dyn} and
 \eqref{Jacobi_dyn_int}. Then we proceed to prove Theorem 1.

\subsection{Additional details about the Weyl function}

 Consider two solutions $P,$ $Q$ of the equation
\begin{equation}
\label{Difer_eqn}
a_{n-1}\psi_{n-1}+b_n\psi_n+a_{n}\psi_{n+1}=\lambda \psi_n, \quad n\geqslant 2,
\end{equation}
satisfying the initial data
\begin{equation*}
P_1(\lambda)=1,\,\,P_2(\lambda)=\frac{\lambda-b_1}{a_1},\quad Q_1(\lambda)=0,\,\,
Q_2(\lambda)=\frac1{a_1}.
\end{equation*}
Thus $P_n(\lambda)$ is a polynomial of degree $n-1$, and
 $Q_n(\lambda)$ is a polynomial of degree $n-2$.

In the case of finite Jacobi matrix $H^N$ we introduce $\Phi^+(\lambda)$ to
be a solution to (\ref{Difer_eqn}) fixed by conditions ``at
the right end'':
\begin{equation*}
\Phi^+_N(\lambda)=1,\quad \Phi^+_{N+1}(\lambda)=0.
\end{equation*}
In the case of semi-infinite $H$,  in order to introduce the solution $\Phi^+(\lambda)$ for
$\lambda\notin \mathbb{R}$, we extend the equation \eqref{Difer_eqn} to $n=1$ using the same additional parameter
$a_0=1$ as before, and define
\begin{equation}
\label{L_2} \Phi^+_0(\lambda)=1,\quad \sum_{n=0}^\infty
|\Phi^+_n(\lambda)|<\infty.
\end{equation}
 Since entries of the matrix are bounded, it is in the limit point case \cite[Chapter 1]{Ahiez}, and hence there exists only one such solution for $\lambda\notin \mathbb{R}$.

The  solution $\Phi^+$ is expressed in terms of $P$ and $Q$ as (cf. \cite{SG}, \cite[Chapter 7]{Berez}, \cite[Chapters 1,2]{Ahiez})
\begin{equation}
\label{L_2sol}
Q_n(\lambda)+m(\lambda)P_n(\lambda)= -\Phi^+_n(\lambda)=-\frac{\Phi^+_n(\lambda)}{\Phi^+_0(\lambda)}.
\end{equation}
 Then for $N<\infty$
\begin{equation}
\label{M_int_def}
m^N(\lambda)=-\frac{Q_{N+1}(\lambda)}{P_{N+1}(\lambda)}=-\frac{\Phi^+_1(\lambda)}{\Phi^+_0(\lambda)},
\end{equation}
and for $N=\infty$
\begin{equation*}
m(\lambda)=-\Phi^+_1(\lambda).
\end{equation*}
Since  the limit point case holds, we have $m^N(\lambda)
\underset{N\to\infty}\longrightarrow m(\lambda) $ uniformly in all bounded
domains in  the upper and  the lower half-planes.

\subsection{Discrete Fourier transformation.}

Consider two solutions of the equation
\begin{equation}
\label{Difer_eqn0} \psi_{n-1}+\psi_{n+1}=\lambda \psi_n,
\end{equation}
satisfying the initial conditions:
\begin{equation*}
P_0=0,\,\,P_1=1,\quad Q_0=-1,\,\, Q_1=0.
\end{equation*}
Clearly, one has:
\begin{equation*}
Q_{n+1}(\lambda)=P_{n}(\lambda).
\end{equation*}
On introducing the notation $U_n(\lambda)=P_{n+1}(2\lambda)$, we
see that $U$ satisfies
\begin{equation}
\label{Difer_eqnU} U_{n-1}(\lambda)+U_{n+1}(\lambda)=2\lambda
U_n(\lambda),\quad U_0=1,\,\,\ U_1=2\lambda.
\end{equation}
So $U_n$ are Chebyshev polynomials of the second kind, and for them the following representation holds:
\begin{eqnarray*}
U_n(\lambda)=\frac{\left(\lambda+\sqrt{\lambda^2-1}\right)^{n+1}-\left(\lambda-\sqrt{\lambda^2-1}\right)^{n+1}}
{2\sqrt{\lambda^2-1}}\\=
\sum_{k=0}^{\left[\frac{n}{2}\right]}C^{n+1}_{2k+1}\left(\lambda^2-1\right)^k\lambda^{n-2k}.
\end{eqnarray*}
We are looking for the unique $l^2$-solution $S$ (see (\ref{L_2}),
(\ref{L_2sol})):
\begin{equation*}
S_n(\lambda)=Q_n(\lambda)+m_0(\lambda)P_n(\lambda),
\end{equation*}
where $m_0$ is the Weyl function of $H_0$ (the special case of $H$
with  $a_n\equiv1$, $b_n\equiv0$). This function can be
obtained as the limit of Weyl functions $m_0^N$ for the problem
(\ref{Difer_eqn0}) with the Dirichlet condition at $n=N+1$:
\begin{equation}
\label{M0_def}
m_0^N(\lambda)=-\frac{Q_{N+1}(\lambda)}{P_{N+1}(\lambda)}
=\frac{P_N(\lambda)}{P_{N+1}(\lambda)}.
\end{equation}
Consider the new variable $z$ related to $\lambda$ by the equalities
\begin{equation}
\label{Lam_z_conn} \lambda=z+\frac1z, \quad
z=\frac{\lambda-\sqrt{\lambda^2-4}}2=\frac{\lambda-i\sqrt{4-\lambda^2}}2,
\end{equation}
$z\in\mathbb D\cap\mathbb C_-$ for $\lambda\in\mathbb C_+$, where
$\mathbb{D}=\{z\,|\, |z|\leqslant 1\}$. Using the formula for
Chebyshev polynomials we can pass to the limit as
$N\rightarrow+\infty$ and  obtain
$$
m_0(\lambda)=-z.
$$
Equation \eqref{Difer_eqn0} has two solutions, $z^n$ and $z^{-n}$.
Since $S\in l^2(\mathbb N)$ and $S_1(\lambda)=m_0(\lambda)=-z$, we
get
$$
S_n(\lambda)=-z^n.
$$
Let $\chi_{(a,b)}(\lambda)$ be the characteristic function of the
interval $(a,b)$. The spectral measure of the unperturbed operator
$H_0$ corresponding to (\ref{Difer_eqn0}) is
\begin{equation*}
d\rho(\lambda)=\chi_{(-2,2)}(\lambda)\frac{\sqrt{4-\lambda^2}}{2\pi}\,d\lambda.
\end{equation*}
Then the Fourier transformation $F: l^2(\mathbb{N})\mapsto
L_2((-2,2),\rho)$ acts by the rule: for $(v_n)\in
l^2(\mathbb{N})$:
\begin{equation*}
F(v)(\lambda)=V(\lambda)=\sum_{n=0}^\infty S_n(\lambda)v_n.
\end{equation*}
The inverse transform is given by:
\begin{equation*}
v_n=\int_{-2}^2V(\lambda)S_n(\lambda)\,d\rho(\lambda).
\end{equation*}

\subsection{Relationship between response vectors and Weyl functions}
 Now we have everything that is necessary to prove Theorem 1.

\begin{proof}[ Proof of Theorem 1.]
In (\ref{Jacobi_dyn}), (\ref{Jacobi_dyn_int}) we take the special
controls: $f=\delta$, $g=\delta$ and go over the Fourier transform
in the variable $t$: for a fixed $n$ we evaluate the sum using the
conditions at $t=0$:
\begin{equation*}
\sum_{t=0}^\infty
S_t(\lambda)\left[u^\delta_{n,t+1}+u^\delta_{n,t-1}-a_nu^\delta_{n+1,t}-a_{n-1}u^\delta_{n-1,t}-b_nu^\delta_{n,t}\right]=0.
\end{equation*}
Changing the order of summation yields:
\begin{equation}
\label{Sum} \sum_{t=0}^\infty
u^\delta_{n,t}\left[S_{t-1}(\lambda)+S_{t+1}(\lambda)\right]-
S_t(\lambda)\left[a_nu^\delta_{n+1,t}+a_{n-1}u^\delta_{n-1,t}+b_nu^\delta_{n,t}\right]=0.
\end{equation}
Introducing the notation
\begin{equation}
\label{U_hat} \widehat u_n(\lambda):=\sum_{t=0}^\infty
S_t(\lambda)u^\delta_{n,t}
\end{equation}
and assuming that $S$ satisfies (\ref{Difer_eqnU}), we deduce from
(\ref{Sum}) that $\widehat u$ satisfies
\begin{equation}
\label{Syst_lambda_infinit} \left\{
\begin{array}l
a_{n-1}\widehat u_{n-1}(\lambda)+a_n\widehat u_{n+1}(\lambda)+b_n\widehat
u_{n}(\lambda)=\lambda\widehat u_n(\lambda),\\
\widehat u_0(\lambda)=-1.
\end{array}
\right.
\end{equation}
Similarly, introducing
\begin{equation}
\label{V_hat}
\widehat v_n(\lambda):=\sum_{t=0}^\infty S_t(\lambda)v^{\delta}_{n,t}
\end{equation}
one can check that $\widehat v$ satisfies
\begin{equation}
\label{Syst_lambda_finit} \left\{
\begin{array}l
a_n\widehat v_{n+1}(\lambda)+a_{n-1}\widehat v_{n-1}(\lambda)+b_n\widehat
v_{n}(\lambda)=\lambda\widehat v_n(\lambda),\\
\widehat v_0(\lambda)=-1, \,\, \widehat v_{N+1}=0.
\end{array}
\right.
\end{equation}
Then by (\ref{M_int_def}) we immediately obtain the representation
for the Weyl function in the finite case:
\begin{equation}
\label{MN_repr}
m^N(\lambda)=\widehat v_1(\lambda)=\sum_{t=0}^\infty
S_t(\lambda)r^N_t.
\end{equation}
As a consequence of finiteness of the signal propagation speed in the
systems (\ref{Jacobi_dyn}), (\ref{Jacobi_dyn_int}) one has local
dependance of the response operator (response vector) on the
coefficients $\{a_n,b_n\}$ (see \cite{MM,MM1}). The latter, in
particular, yields that $r_t=r^N_t$ for $t=0,1,\ldots,2N-1$,
which, in turn, leads to the convergence
\begin{equation}
\label{S_conv} \sum_{t=0}^\infty
S_t(\lambda)r^N_t\underset{N\to\infty}\longrightarrow \sum_{t=0}^\infty
S_t(\lambda)r_t,
\end{equation}
which takes place in the region which will be specified below.
Since $H$ is in the limit point case,
Weyl functions $m^N$ of finite-dimensional operators $H^N$ converge to $m$ uniformly in bounded domains
in $\mathbb{C}_+$ and $\mathbb{C}_-$ as $N\to\infty$, so we deduce
that
\begin{equation}
\label{M_repr} m(\lambda)=\sum_{t=0}^\infty
S_t(\lambda)r_t=\widehat u_1(\lambda)=-\sum_{t=0}^\infty z^tr_t.
\end{equation}
Passing from dynamical systems with discrete time
(\ref{Jacobi_dyn}), (\ref{Jacobi_dyn_int}) to systems
(\ref{Syst_lambda_infinit}), (\ref{Syst_lambda_finit}) with  the
parameter $\lambda$ will be justified as soon as we show that sums
in (\ref{U_hat}) and (\ref{V_hat}) converge. To this
 end we need estimates on $|u^\delta_{n,t}|$,
$|v^\delta_{n,t}|$. We treat the semi-infinite case, the arguments
for the estimate for $|v^\delta_{n,t}|$ are the same. Introducing
the notation
\begin{equation*}
M_t:={\rm max}_{1\leqslant n\leqslant
t}\left\{|u^\delta_{n,t}|,|u^\delta_{n,t-1}|\right\},
\end{equation*}
from the difference equation (the first line) of the system
(\ref{Jacobi_dyn}) we have the following estimate:
\begin{equation*}
|u^\delta_{n,t+1}|\leqslant (3B+1)M_t.
\end{equation*}
From this relation we obtain that
\begin{equation*}
M_{t+1}\leqslant (3B+1)M_t,\quad M_0=1.
\end{equation*}
Thus $M_t$ is bounded by the following:
\begin{equation*}
M_{t}\leqslant (3B+1)^t.
\end{equation*}
The above estimate implies that the sum in (\ref{U_hat})
converges, provided
\begin{equation*}
\left|z\left(\lambda\right)\right|^t(3B+1)^t<1,
\end{equation*}
or
\begin{equation}
\label{xi_est} \left|z(\lambda)\right|<\frac{1}{3B+1}.
\end{equation}
Denote $R:=3B+1$. Then (\ref{xi_est}) is valid in the region
$D\subset \mathbb{C}_+$ specified in (\ref{D_reg}), which finishes
the proof of Theorem \ref{Theor_main}.
\end{proof}

\begin{remark}
In $z$ variable the formulas (\ref{M_inf}), (\ref{M_fin}) gives a
Taylor expansions of Weyl functions at zero, these expansions hold
in the region $\mathbb{C}_-\cap
\{\left|z\right|<\frac{1}{3B+1}\}$.
\end{remark}

\noindent{\bf Acknowledgments}

The research of Victor Mikhaylov was supported in part by RFBR
17-01-00529. Alexandr Mikhaylov was supported by RFBR 17-01-00099;
A. S. Mikhaylov and V. S. Mikhaylov were partly supported by RFBR
18-01-00269 and by VW Foundation program ``Modeling, Analysis, and
Approximation Theory toward application in tomography and inverse
problems.'' S.  A. Simonov was supported by grants RFBR
17-01-00529, RFBR 16-01-00443A and RFBR 16-01-00635A.


\begin{thebibliography}{99}


\bibitem{ABu} {S. A. Avdonin, A. S. Bulanova}. \textit{Boundary
control approach to the spectral estimation problem. The case of
multiple poles}, Math. Contr. Sign. Syst. 22, no. 3, 245--265,
2011.


\bibitem{ABuN} {S. A. Avdonin, A. S. Bulanova, D. J.
Nicolsky}. \textit{Boundary control approach to the spectral
estimation problem.  The case of simple poles}, Sampling Theory in
Signal and Image Processing, 8, no. 3, 225--248, 2009.


\bibitem{AM}
{S. A. Avdonin, V. S. Mikhaylov}. \textit{The boundary control
approach to inverse spectral theory,} Inverse Problems, 26, no. 4,
045009, 19 pp, 2010.

\bibitem{AMR}
{S. A. Avdonin, V. S. Mikhaylov, A. V. Rybkin}. \textit{The
boundary control approach to the Titchmarsh-Weyl $m-$function},
Comm. Math. Phys. 275, no. 3, 791--803, 2007.

\bibitem{Ahiez}
{N. I. Akhiezer}. \textit{The classical moment problem and some
related questions in analysis.} Oliver and Boyd, 1965.

\bibitem{A}
{F. V. Atkinson}. \textit{Discrete and continuous boundary
problems.} Acad. Press, 1964.

\bibitem{B07}
{M.I. Belishev}. \textit{Recent progress in the boundary control
method}, Inverse Problems, 23, no. 5, R1--R67, 2007.

\bibitem{B17}
{M.I.Belishev.} \textit{Boundary control and tomography of
Riemannian manifolds (the BC-method).} Uspekhi Matem. Nauk, 72,
no. 4, 3-66, 2017, (in Russian).

\bibitem{B01JII}
{M. I. Belishev}. \textit{On relation between spectral and
dynamical inverse data}, J. Inv. Ill-posed problems. 9, no. 6,
647--665, 2001.

\bibitem{BM}
{M. I. Belishev, V. S. Mikhaylov}. \textit{Unified approach to
classical equations of inverse problem theory.} {Journal of
Inverse and Ill-Posed Problems}, 20, no. 4, 461--488, 2012.

\bibitem{Berez}
{Ju. M. Berezanskii}. \textit{Expansions in eigenfunctions of selfadjoint operators}
American Mathematical Society, Vol. {\bf 17}, 1968.

\bibitem{I}
{V. Isakov}. \textit{Inverse problems for partial differential
equations.} Appl. Math. Studies, Springer, v. 127, 1998.

\bibitem{SG}
{F. Gesztesy, B. Simon}. \textit{$m-$functions and inverse
spectral analisys for finite and semi-infinite Jacobi matrices},
J. d'Analyse Math. 73, 267-297, 1997.

\bibitem{KKL}
{A. P. Kachalov, Y. V. Kurylev, M. Lassas}. \textit{Inverse
Boundary Spectral Problems.,} Chapman\&Hall, 2001.

\bibitem{MM}
{A. S. Mikhaylov, V. S Mikhaylov}. \textit{Dynamic inverse problem
for the discrete Schr\"odinger operator.} Nanosystems: Physics,
Chemistry, Mathematics, 7, no. 5, 842-854, 2016.

\bibitem{MM1}
{A. S. Mikhaylov, V. S Mikhaylov}. \textit{Dynamic inverse problem
for Jacobi matrices.} \textit{https://arxiv.org/abs/1704.02481}.

\bibitem{MM2}
{A. S. Mikhaylov, V. S Mikhaylov}. \textit{Relationship between
different types of inverse data for the one-dimensional
Schr\"odinger operator on a half-line.} \textit{Zap. Nauchn.
Semin. POMI}, 451, 134-155, 2016, in J. Math. Sci. (N.Y.), 226,
no. 6, 779-–794, 2017.

\bibitem{MM3}
{A.S. Mikhaylov, V.S Mikhaylov}. \textit{Boundary Control method
and de Branges spaces. Schr\"odinger operator, Dirac system,
discrete Schr\"odinger operator.} Journal of Mathematical Analysis
and Applications, 460, no. 2, 927-953, 2018.

\bibitem{S}
{B. Simon}. \textit{The classical moment problem as a self-adjoint
finite difference operator.},  Advances in Math., 137, 82-203,
1998.

\bibitem{BS1}
B. Simon. \textit{A new approach to inverse spectral theory, I.
Fundamental formalism.,} Annals of Mathematics, 150, 1029--1057,
1999.

\bibitem{T}
G. Teschl. \textit{Jacobi Operators and Completely Integrable
Nonlinear Lattices.} American Mathematical Society, Vol. {\bf 72},
2000.

\end{thebibliography}
\end{document}